# Biased decoy-state reference-frame-independent quantum key distribution

Jian-Rong Zhu,[1,2,#] Chun-Mei Zhang,[1,2,#] Guang-Can Guo,[1,2,3] Qin Wang [1,2,3,*]

[1] Institute of Signal Processing Transmission, Nanjing University of Posts and Telecommunications, Nanjing, 210003, China

[2] Key Lab of Broadband Wireless Communication and Sensor Network Technology, Nanjing University of Posts and Telecommunications, Ministry of Education, Nanjing, 210003, China

3. Key Laboratory of Quantum Information, CAS, University of Science and Technology of China, Hefei 230026, China

*Corresponding author: qinw@njupt.edu.cn

## Abstract

Without actively aligning the reference frames, the reference-frame-independent quantum key distribution (RFI-QKD) can generate secure keys even when the reference frames drift slowly. Here, we propose a new scheme on the decoy-state RFI-QKD, where the basis assignments of two legitimate parties are biased, reducing the fraction of mismatched-basis data. When accounting for statistical fluctuations under different misalignments of reference frames, we investigate the performance of biased decoy-state RFI-QKD with full parameter optimization. We also make a clear comparison between the new scheme with other practical schemes, e.g., unbiased decoy-state RFI-QKD and biased decoy-state BB84-QKD. Simulation results demonstrate that the new proposed biased decoy-state RFI-QKD can significantly improve the performance (secret key rate and secure transmission distance) of practical QKD systems.

## Introduction

Quantum key distribution (QKD) can provide a feasible approach for two remote peers, Alice and Bob, to share unconditionally secure keys even in the presence of an eavesdropper, Eve [1-3]. Since Bennett and Brassard proposed the first well known BB84 QKD protocol in 1984 [4], a lot of achievements have been made both in theories [5-7] and experiments [8-13]. In most of reported QKD experiments, the reference frames should be actively aligned between Alice and Bob to assure the regular running of practical QKD systems. However, the alignment of reference frames will take up considerable long time overhead which could generate secret keys otherwise, and also makes QKD systems more complicated which may be deployed by Eve [14-15].

Thanks to the reference-frame-independent quantum key distribution (RFI-QKD) protocol proposed by Laing et al. [16], Alice and Bob may share secret keys without the alignment of reference frames. Moreover, due to the unavailability of ideal single-photon sources, phase-randomized weak coherent sources which emit multi-photon states with nonzero probabilities are widely used in practical QKD systems. However, Eve can easily hack those multi-photon components by performing the photon-number-splitting (PNS) attack without Alice and Bob's awareness [17-19]. To combat the PNS attack, the decoy-state method [20-25] was adopted in practical RFI-QKD systems. Up to date, there have been a few theoretical [26-27] and experimental [28-31] work on RFI-QKD. However, there exists shortcomings in each of them, e.g., the decoy-state method was not employed in Refs. [27,30,31], the issue of statistical fluctuations was not taken into account in Refs. [26,29], and it used fixed basis choice and intensities in [28].

In this paper, we propose the high efficient decoy-state RFI-QKD scheme, where Alice and Bob's basis choices are biased to increase the fraction of matched-basis data. When accounting for statistical fluctuations and different misalignments of reference frames, we investigate our protocol with full parameter optimization [32]. Comparisons between this new proposed protocol and either unbiased decoy-state RFI-QKD or biased decoy-state BB84-QKD are made. Simulation results show that the new proposed biased decoy-state RFI-QKD can significantly improve the performance of practical QKD systems. The structure of the paper is as follows: In Sec II, we demonstrate biased decoy-state RFI-QKD protocol. In Sec III, we show corresponding simulation results. Finally, conclusions are given.

## Biased decoy-state RFI-QKD

In perfect QKD systems, the reference frames $X$, $Y$ and $Z$ are well aligned, where $X = |0\rangle\langle 1| + |1\rangle\langle 0|, Y = -i|0\rangle\langle 1| + i|1\rangle\langle 0|$ and $Z = |0\rangle\langle 0| - |1\rangle\langle 1|$. However, in most practical QKD systems, only one reference frame can be well aligned [16]. In RFI-QKD, Alice randomly prepares quantum states according to her bases $X_A$, $Y_A$ and $Z_A$, and sends them to Bob. Once received the states from Alice, Bob measures those states according to his bases $X_B$, $Y_B$ and $Z_B$. In this scheme, $Z$ basis is stable which is used to distill secret keys. $X$ and $Y$ bases are allowed to drift slowly which are used to estimate Eve's information. In other words, they follow:

$$Z_B = Z_A \tag{1}$$

$$X_B = \cos\beta X_A + \sin\beta Y_A, \tag{2}$$

$$Y_B = \cos\beta Y_A - \sin\beta X_A, \tag{3}$$

where $\beta$ is the misalignment of Alice and Bob's local $X$ and $Y$ bases.

Here, we propose the biased decoy-state RFI-QKD, and we concentrate on the widely used 3-intensity decoy-state scheme, which employs one signal state (with an intensity of $\mu$), one weak decoy state (with an intensity of $\nu$) and one vacuum state (with an intensity of 0). Alice randomly modulates pulses into three different intensities $u$, $v$ and $0$ with probabilities $P_\mu$, $P_\nu$ and $1 - P_\mu - P_\nu$, respectively. For pulses with intensity $\mu$ ($\nu$), she prepares states in her $Z_A$ basis with conditional probability $P_{Z_A|\mu}$ ($P_{Z_A|\nu}$), $X_A$ basis with probability $P_{X_A|\mu}$ ($P_{X_A|\nu}$) and $Y_A$ basis with probability $1 - P_{Z_A|\mu} - P_{X_A|\mu}$ ($1 - P_{Z_A|\nu} - P_{X_A|\nu}$). For vacuum states, Alice does not set any basis. Furthermore,

Bob measures the incoming pulses by randomly choosing bases $Z_B$, $X_B$ and $Y_B$ with probabilities $P_{Z_B}$, $P_{X_B}$ and $1 - P_{Z_B} - P_{X_B}$.

When taking statistical fluctuations into account, the total number of pulses sent by Alice can be denoted as $N$. Following the statistical fluctuation analysis in Refs. [33,34], the overall gain and quantum-bit error of states that Alice sends in $\xi_A$ ($\xi_A \in \{X_A, Y_A, Z_A\}$) basis and Bob measures in $\xi_B$ ($\xi_B \in \{X_B, Y_B, Z_B\}$) basis can be estimated by:

$$Q_{\xi_A \xi_B}^{\mu,U} = Q_{\xi_A \xi_B}^{\mu} \left(1 + \frac{\gamma}{\sqrt{N P_\mu P_{\xi_A|\mu} P_{\xi_B} Q_{\xi_A \xi_B}^{\mu}}}\right), \tag{4}$$

$$Q_{\xi_A \xi_B}^{\nu,L} = Q_{\xi_A \xi_B}^{\nu} \left(1 - \frac{\gamma}{\sqrt{N P_\nu P_{\xi_A|\nu} P_{\xi_B} Q_{\xi_A \xi_B}^{\nu}}}\right), \tag{5}$$

$$E_{\xi_A \xi_B}^{\nu,U} Q_{\xi_A \xi_B}^{\nu,U} = E_{\xi_A \xi_B}^{\nu} Q_{\xi_A \xi_B}^{\nu} \left(1 + \frac{\gamma}{\sqrt{N P_\nu P_{\xi_A|\nu} P_{\xi_B} E_{\xi_A \xi_B}^{\nu} Q_{\xi_A \xi_B}^{\nu}}}\right), \tag{6}$$

where superscripts $\mu$ and $\nu$ represent signal states and decoy states respectively, superscripts $U$ and $L$ denote the upper and lower bound of experimental results, and $\gamma$ is the number of standard deviations which is directly related to the failure probability of the security analysis. Similarly, we can also get:

$$Q_0^L = Q_0 \left(1 - \frac{\gamma}{\sqrt{N(1 - P_\mu - P_\nu) Q_0}}\right), \tag{7}$$

$$Q_0^U = Q_0 \left(1 + \frac{\gamma}{\sqrt{N(1 - P_\mu - P_\nu) Q_0}}\right). \tag{8}$$

With the formulae above and decoy-state analytical method [35], we can estimate the lower bound of the yield and the upper bound of the error rate of single-photon contributions in $Z_A Z_B$, denoted as $Y_{Z_A Z_B}^{1,L}$ and $e_{Z_A Z_B}^{1,U}$, respectively. Similarly, the upper bound of the error rate of single-photon contributions in $X_A X_B$, $X_A Y_B$, $Y_A X_B$ and $Y_A Y_B$ can also be estimated, each denoted as $e_{X_A X_B}^{1,U}$,

$e_{X_AY_B}^{1,U}$, $e_{Y_AX_B}^{1,U}$ and $e_{Y_AY_B}^{1,U}$. Then the intermediate quantity $C$ is given by:

$$C = \left(1-2e_{X_AX_B}^{1,U}\right)^2 + \left(1-2e_{X_AY_B}^{1,U}\right)^2 + \left(1-2e_{Y_AX_B}^{1,U}\right)^2 + \left(1-2e_{Y_AY_B}^{1,U}\right)^2. \tag{9}$$

And Eve' information $I_E$ is:

$$I_E = \left(1-e_{Z_AZ_B}^{1,U}\right)H\left(\frac{1+\phi}{2}\right) + e_{Z_AZ_B}^{1,U}H\left(\frac{1+\varphi}{2}\right), \tag{10}$$

where $H(x)$ is the binary Shannon information function, given by $H(x) = -x\log_2(x) - (1-x)\log_2(1-x)$, and

$$\phi = \min\left[\frac{1}{1-e_{Z_AZ_B}^{1,U}}\sqrt{C/2}, 1\right], \tag{11}$$

$$\varphi = \sqrt{C/2 - \left(1-e_{Z_AZ_B}^{1,U}\right)^2 \phi^2} \Big/ e_{Z_AZ_B}^{1,U}. \tag{12}$$

According to the GLLP [36] security analysis, the final key generation rate can be calculated as:

$$R \geq P_\mu P_{Z|\mu} P_Z \left\{-fQ_{Z_AZ_B}^\mu H\left(E_{Z_AZ_B}^\mu\right) + \mu e^{-\mu} Y_{Z_AZ_B}^{1,L}\left[1 - H\left(e_{Z_AZ_B}^{1,U}\right)\right]\right\}, \tag{13}$$

where $Q_{Z_AZ_B}^\mu$ and $E_{Z_AZ_B}^\mu$ represent the average counting rate and the quantum-bit error-rate of signal states in $Z_AZ_B$ basis, respectively. $f$ is a factor to evaluate the reconciliation efficiency in key reconciliation phase.

## Numerical simulation

In order to acquire better performance with respect to key generation rate and secure transmission distance, full parameter optimization is carried out in this biased decoy-state RFI-QKD with statistical fluctuations, where ten parameters, including $\mu$, $\nu$, $P_\mu$, $P_\nu$, $P_{Z_A|\mu}$, $P_{Z_A|\nu}$, $P_{X_A|\mu}$, $P_{X_A|\nu}$, $P_{Z_B}$ and $P_{X_B}$, need to be optimized. The experimental parameters [37] listed in Table. 1 are employed in all the numerical simulation results presented in this section.

| $\eta$ | $Y_0$ | $e_d$ | $\alpha$ | $f$ | $\gamma$ |
|---|---|---|---|---|---|
| 14.5% | $3.0 \times 10^{-6}$ | 1.5% | 0.2dB/km | 1.16 | 5 |

Table. 1. List of practical parameters for numerical simulations. $\eta$ is the detection efficiency of the single-photon detector, $Y_0$ is the dark count rate of the single-photon detector, $e_d$ is the probability of a photon arriving at the erroneous single-photon detector, $\alpha$ is the loss coefficient of optical fiber, $f$ is the key reconciliation efficiency in reconciliation phase, $\gamma$ is the confidence interval for statistical fluctuation with a failure probability of $5.73 \times 10^{-7}$ [33].

Considering statistical fluctuations, we investigate the biased decoy-state RFI-QKD at $\beta=0°, 10°, 20°$ with full parameter optimization. For the sake of contrast, we also investigate the performance of unbiased decoy-state RFI-QKD with reasonable parameters, i.e., $\nu = 0.1$, $P_\mu = P_\nu = 1/3$, $P_{Z_A|\mu} = P_{X_A|\mu} = 1/3$, $P_{Z_A|\nu} = P_{X_A|\nu} = 1/3$, $P_{Z_B} = P_{X_B} = 1/3$, and optimized $\mu$ at each distance [28]. All the key rates are simulated with a reasonable data size at $N=10^{11}$ [32].

As shown in Fig.1, the key generation rates of the biased decoy-state RFI-QKD at different misalignments of reference frames have been improved with more than one order of magnitude, compared with those of unbiased decoy-state RFI-QKD. In addition, it is easy to see that the performance of biased decoy-state RFI-QKD will be depressed by the increasing of the rotation angle of the reference frames. e.g, the key rate of biased decoy-state RFI-QKD with $\beta=20°$ is lower than that of $\beta=10°$. Despite with biggish rotation angle, the key rate of biased decoy-state RFI-QKD with $\beta=20°$ is still higher than that of unbiased decoy-state RFI-QKD with $\beta=0°$ when the distance is shorter than 140 km. Therefore, it is significant to adopt our biased decoy-state RFI-QKD scheme which can remarkably improve the performance of practical RFI-QKD systems.

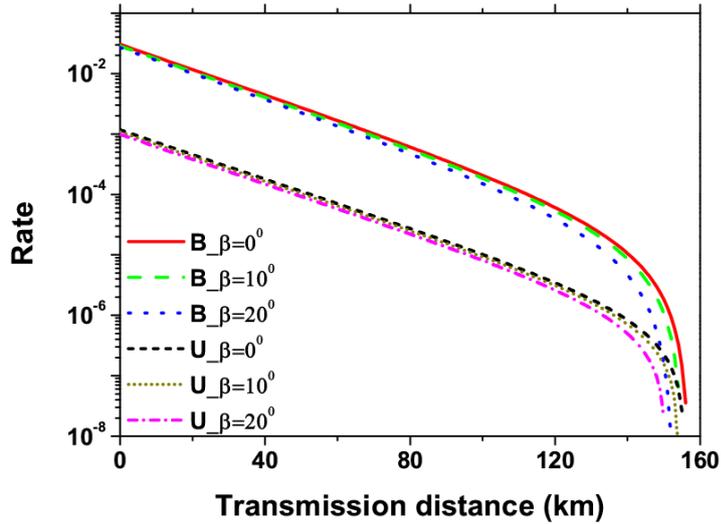

Fig. 1. Comparison of the key generation rates between the full optimized biased and unbiased decoy-state RFI-QKD at different misalignments of reference frames. The curves from top to bottom are the secret key rates of biased decoy-state RFI-QKD with $\beta=0°, 10°, 20°$, and unbiased decoy-state RFI-QKD with $\beta=0°, 10°, 20°$, respectively.

By adopting full parameter optimization, we also make a comparison between the biased decoy-state RFI-QKD and biased quantum key distribution with BB84 protocol (BB84-QKD) at different misalignments of reference frames with $N=10^{11}$. Corresponding simulation results are shown in Fig.2. Generally, the reference frames between Alice and Bob in biased decoy-state BB84-QKD should be actively aligned, which will obviously take up considerable long time overhead. And the time of alignments is difficult to evaluate. For simplicity, we consider biased BB84-QKD suffers the same misalignments as biased RFI-QKD. As shown in Fig.2, the key generation rates of biased decoy-state RFI-QKD decline slowly with the increase of rotation angles, compared with those of biased decoy-state QKD.

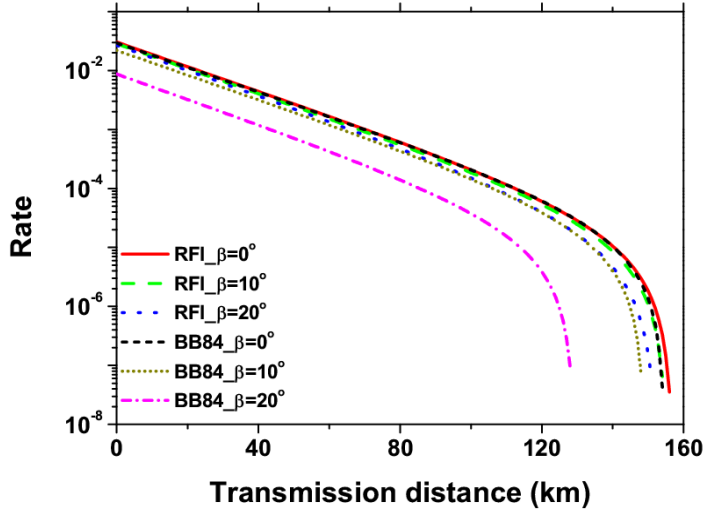

Fig. 2. Comparison of the key generation rates between the biased decoy-state RFI-QKD and BB84-QKD at different misalignments of reference frames. The solid curve, dash curve and dot curve refer to RFI-QKD with $\beta=0°, 10°, 20°$, respectively. The short dash curve, short dot curve and short dash dot curve refer to BB84-QKD with $\beta=0°, 10°, 20°$, respectively.

Furthermore, the key rates of biased decoy-state RFI-QKD with various total numbers of pulses from $10^9$ to $10^{13}$ at a fixed distance of 100 km are displayed in Fig.3. From this figure it is obvious that the key generation rates of biased decoy-state RFI-QKD decline less than one order of magnitude when data size $N$ reduces from $10^{13}$ to $10^9$. Even though with short communication time and small data size, biased decoy-state RFI-QKD can still generate considerable secret keys. Hence, it is efficient to adopt the biased decoy-state RFI-QKD under the situation when the reference frames drift slowly.

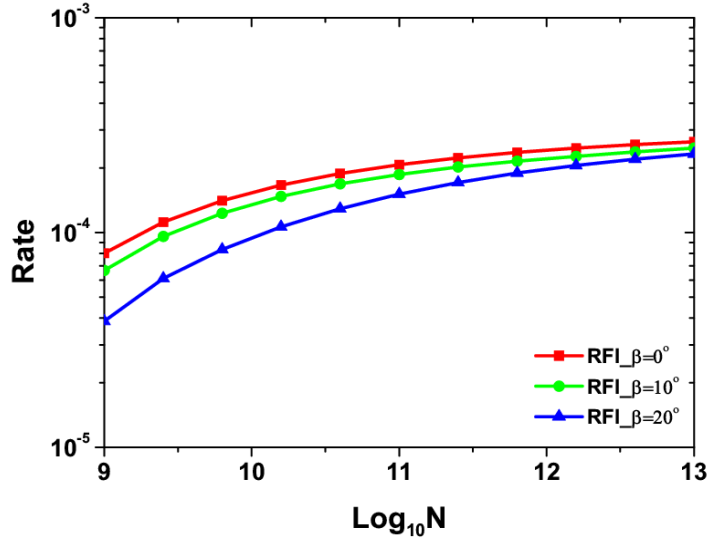

Fig. 3. The optimized key generation rates (per pulse) versus different total numbers of pulses N for biased decoy-state RFI-QKD at the distance of 100 km with different rotation angles. Curves from top to bottom refer to biased decoy-state RFI-QKD with $\beta=0°, 10°, 20°$, respectively.

## Conclusions

In summary, we have proposed a biased decoy-state RFI-QKD scheme, where Alice and Bob's basis choices are set biased to increase the fraction of matched-basis data. We carry out the full parameter optimization method to optimize the achievable key generation rate and transmission distance. Our simulation results demonstrate that the key generation rate of full optimized biased decoy-state RFI-QKD can be enhanced by more than one order of magnitude compared to those of unbiased decoy-state RFI-QKDs. Moreover, considering the influence of statistical fluctuations and different misalignments of reference frames, the biased decoy-state RFI-QKD is more efficient and practicable than those biased decoy-state BB84-QKDs. We hope that this work could provide a valuable reference for practical implementations of decoy-state RFI-QKDs.

We gratefully acknowledge the financial support from the National Natural Science Foundation of China through Grants No. 11274178, No. 61475197 and No. 61590932, the Natural Science Foundation of the Jiangsu Higher Education Institutions through Grant No. 15KJA120002, the Outstanding Youth Project of Jiangsu Province through Grant No. BK20150039, and the Priority Academic Program Development of Jiangsu Higher Education Institutions through Grant No. YX002001.

# These authors contributed equally to this work.